\documentclass[prd,twocolumn,amsmath,amsfonts,amssymb,showpacs,nofootinbib]{revtex4}
\usepackage{epsfig,graphicx,bm,color}
\begin{document}
\title{Complementarity of direct dark matter detection\\ and indirect detection through gamma-rays}
\author{Lars Bergstr\"om}
\email{lbe@fysik.su.se}
\affiliation{The Oskar Klein Centre for Cosmoparticle Physics, Department of Physics, Stockholm University, AlbaNova, SE-106 91 Stockholm, Sweden}
\author{Torsten Bringmann}
\email{torsten.bringmann@desy.de}
\affiliation{{II.} Institute for Theoretical Physics, University of Hamburg, 
Luruper Chausse 149, D-22761 Hamburg, Germany}
\author{Joakim Edsj\"o}
\email{edsjo@fysik.su.se}
\affiliation{{T}he Oskar Klein Centre for Cosmoparticle Physics, Department of Physics, Stockholm University, AlbaNova, SE-106 91 Stockholm, Sweden}

\date{February 23, 2011}


\begin{abstract}
We show, by using an extensive sample of viable supersymmetric models as templates, that indirect detection of dark matter through gamma rays may have a large potential for identifying the nature of dark matter. This is in particular true also for models that give  too weak dark matter-nucleon scattering cross sections to be probed by present and planned direct detection experiments. Also models with a mass scale too high to be accessible at CERN's LHC accelerator may show up in next-generation imaging Cherenkov telescope arrays. Based on our our findings, we therefore suggest to view indirect searches as genuine particle physics experiments, complementing other strategies to probe so far unknown regions in the parameter space of e.g. supersymmetric models, and propose a new approach that would make use of  telescopes \emph{dedicated}  for dark matter searches. As a concrete example for the potential of such an approach, we consider an array of imaging air Cherenkov telescopes, the Dark Matter Array (DMA), and show that such an experiment could extend present-day limits by several orders of magnitude, reaching a large class of models that would remain undetected in both direct detection experiments and searches at the LHC. In addition, in a sizable part of the parameter space, signals from more than one type of dark matter detection experiment would be possible, something that may eventually be necessary in order to identify the dark matter candidate.
\end{abstract}

\maketitle

\newcommand{\ga}{\gamma}
\newcommand{\be}{\begin{equation}}
\newcommand{\ee}{\end{equation}}
\newcommand{\bea}{\begin{eqnarray}}
\newcommand{\eea}{\end{eqnarray}}
\newcommand{\ds}{{\sf DarkSUSY}}
\newcommand{\py}{{\sf PYTHIA}}
\newcommand{\code}[1]{{\tt #1}}

\hyphenation{}

\section{Introduction}
The problem of the nature of the dark matter in the universe remains unsolved, despite great progress in direct detection \cite{cdmsii,xenon100}, indirect detection through gamma-rays 
\cite{fermidmreview} and neutrinos \cite{icecube}. It remains to be seen whether the third distinct method of obtaining information about possible elementary particles beyond the Standard Model of particle physics, detection at CERN's LHC or Fermilab's Tevatron collider, will give any clue to the solution of the problem.

Whereas there has been steady progress in the limits arising from direct detection methods, there will be a more modest improvement of indirect detection limits as Fermi and IceCube will continue to collect data over the next decade. Also, it is not excluded that the particles making up the dark matter may have a mass entering the TeV range, making also the possibility of progress from particle physics accelerators more dismal. On the other hand, the proposed imaging air Cherenkov telescope  (IACT) array, CTA \cite{cta}, could in principle give a substantial improvement of dark matter limits also at the TeV mass scale. However, CTA will be a multipurpose detector with the design and projected observation time to a large extent driven by the science goals of the physics of extreme objects (active galactic nuclei, supernova remnants, etc), with exposure time on dark matter-related targets probably limited to some 50 hours or so. 

In this paper we point out that, by examining the allowed parameter space of one of the most natural dark matter candidates, 
the neutralino in the Minimal Supersymmetric Extension of the Standard Model (the MSSM), there are some ten orders of magnitudes in cross section that will not be reached even by the most ambitious direct detection experiments presently proposed. We also show the remarkable, and somewhat surprising, fact that indirect detection rates for gamma-ray detection of dark matter annihilation in the galactic halo (or subhalos) are very weakly correlated with direct detection rates. This means that a \emph{dedicated gamma-ray detector for dark matter detection may probe} from an orthogonal direction the parameter space of \emph{viable dark matter models, down to direct detection levels that would never be realistically achievable otherwise}. 

\section{Methods of Dark Matter Detection}

The methods of detection of dark matter (for reviews, see \cite{reviews}) 
can be divided into {\em accelerator} production and detection of missing 
energy (especially at the LHC at CERN, which has started operation in spring 2010 and is currently running at half the nominal energy),
 {\em direct  detection} (of dark matter particles impinging on a 
terrestrial detector, with recent impressive upper limits reported by 
\cite{cdmsii,xenon100}), or {\em indirect detection} of particles
generated by the annihilation of dark matter particles in the various galactic and extragalactic structures \cite{fermidmreview} or in the Sun/Earth \cite{icecube}.
All these methods are indeed complementary \cite{DM_complementarity} -- it is likely that a signal 
from more than one type of  experiment will be needed to fully identify the 
particle making up the dark matter. The field is just entering very 
interesting times, with the LHC soon
hopefully giving useful results and new solid state detectors \cite{cdmsii,cogent} and those using liquid noble gases being expected to further strengthen the already presented interesting limits. 

Among indirect detection experiments, the neutrino detector IceCube \cite{icecube} has presented data that especially for annihilation in the sun is starting to become interesting. For gamma rays 
from dark matter annihilation in the halo, Fermi has delivered bounds that have started to touch the parameter space of viable models, 
 as predicted in \cite{prelaunch}, 
in particular for dwarf spheroidal galaxies
\cite{fermidwarfs} and galaxy clusters \cite{fermiclusters};
 the extragalactic diffuse signal detected by FERMI \cite{fermiextra}, as well as all-sky searches for line-signals \cite{fermilines}, have so far not resulted in very stringent limits.
For indirect detection through antimatter, the satellite PAMELA 
launched a few years ago has revealed surprising data on the ratio between positrons and electrons \cite{pamelaratio}. In fact, taken together with data on the sum of electrons and positrons from FERMI \cite{fermipositrons} and HESS \cite{hesspositrons}, a decent, though admittedly somewhat contrived, fit using a dark matter candidate with very peculiar and \emph{a priori} completely unexpected properties can be performed \cite{bez,strumia}. As there are more conventional explanations (e.g. positrons and electrons generated by supernova remnants \cite{grasso}) for these data, we will not consider such models here (although we note that the type of detector we propose here would be sensitive to the high-energy photon emission expected from these models).
 
One problem with all these discovery methods is that the signal searched for 
may be 
quite weak and, in many cases, greatly dominated by the much larger backgrounds. For indirect detection
through gamma-rays, the situation
may in principle be better, due to (i) the direct propagation from the region of production, without 
significant absorption or scattering; (ii)  
the dependence of the annihilation rate on the square of the dark matter density which 
may
give "hot spots" near density concentrations as those predicted by N-body 
simulations;
(iii) possible characteristic features like gamma-ray lines \cite{lines} or steps \cite{fsr}, 
given by the fact that
no more energy than $m_{\chi}$ per particle can be released in the annihilation of 
two non-relativistic 
dark matter particles (we denote the dark matter particle by $\chi$).
It
was recently realised that there could be other important spectral 
features, and
in particular that internal bremsstrahlung (IB) \cite{ourpaper} from 
charged particles produced
in the annihilations could yield a detectable "bump" near the 
highest energy, thus facilitating detection; such signatures could also be used to discriminate between different dark matter candidates, already with the energy resolution of current detectors \cite{IBextra,Bringmann:2008kj}.

When comparing direct and indirect dark matter detection strategies in general, one should keep in mind that direct detection experiments are single-purpose instruments while gamma-ray observatories, as already pointed out in the introduction, usually are multi-purpose experiments dedicated to the physics of extreme objects, with dark matter often being considered but a side-aspect of the general science goals. Of course, this has consequences for the total sensitivity to dark matter searches since  even for CTA one can probably not realistically expect to allocate more than 5-10\% of the available observation time for dark matter searches. For space-based experiments, on the other hand, the effective area will always be a limiting factor.
A \emph{dedicated} experiment for the indirect detection of dark matter could therefore greatly improve the discovery potential of this type of searches; regarded as a particle physics experiment, it would probe otherwise inaccessible regions of masses and couplings of MSSM models (or, more generally, of models for Weakly Interacting Massive Particles, WIMPs).

We here work out some details for a specific ground based detector, the DMA (Dark Matter Array), which seems to be feasible both from the technical and funding points of view. 
Although we treat some of the suitable targets and detection strategies briefly, we consider this as but the first preliminary idea of such a detector, the details of which certainly have to be elaborated much more in future work. Of course, if the LHC would give hints on the mass region of the dark matter particle, opimization using this information could be performed. A particularly interesting scenario would be if the LHC discovers a light Higgs, consistent with supersymmetry, whereas the mass scale would be outside LHC's reach: as multi-TeV gamma-ray detection is routinely done with present day IACT's, this would indicate a unique window of opportunity for DMA. As we will show, indirect detection with a dedicated experiment like DMA would for many models be superior to direct detection also for more canonical neutralino masses in the 100 GeV range.

\section{MSSM and direct detection}
We will use as our template models several hundred thousand MSSM models produced in large scans of parameter space using the \ds package \cite{darksusy}. In addition, we use more restriced mSUGRA models \cite{msugra}, but we will in our plots add the two model sets. We apply all current bounds and Standard Model parameter estimates from accelerator measurements, and we only keep models that are within 3$\sigma$ of the latest WMAP estimate of the dark matter density \cite{Komatsu:2010fb}. 

\begin{figure}[t!]
 \includegraphics[width=0.95\columnwidth]{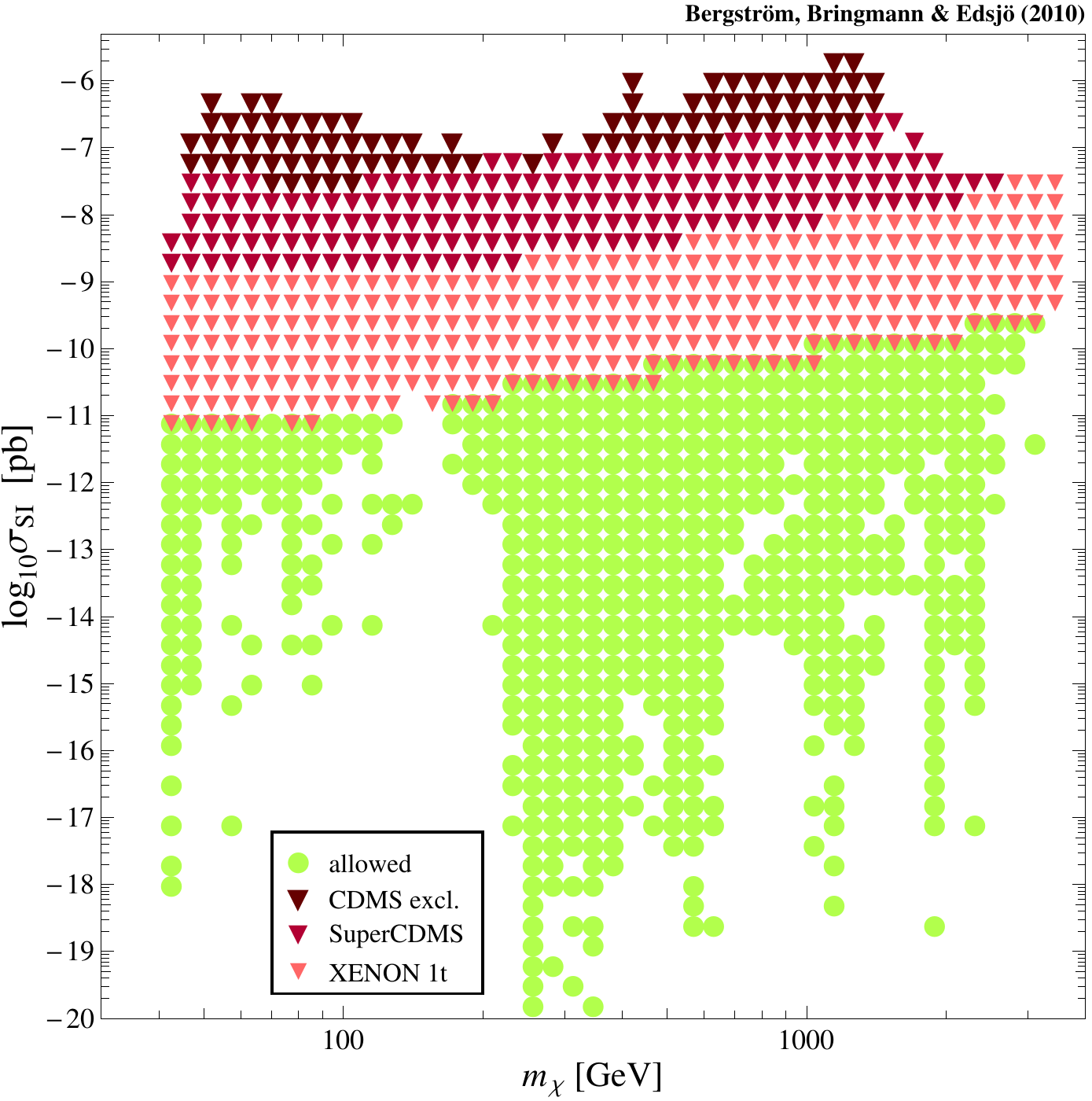}
\caption{The results for the spin independent cross section on protons versus neutralino mass for the sample of MSSM and mSUGRA models found in our scan. Shown are the current exclusion limits from the CDMS experiment \cite{cdms} as well as projected limits for the upgrade to SuperCDMS \cite{supercdms} and the proposed 1t final stage of the XENON experiment \cite{xenon1t}.}
 \label{figDD}
\end{figure}

The set of models we use is compatible, e.g., with that generally shown in plots of direct detection sensitivity \cite{berkeleydmplotter}. However, usually only models that predict a detection rate larger than, say, $10^{-10}$ pb are shown (as such cross sections are of relevant 
size for detectors of the coming decade). In reality, the range of predictions is {\em much} larger. This is seen in Fig.~\ref{figDD}, where the whole range of masses and spin-independent cross sections on protons is shown in a log-log plot. For comparison, we also include the current exclusion limits from the CDMS experiment \cite{cdms}, together  with projected limits for its upgrade to SuperCDMS \cite{supercdms}, as well as the proposed 1t final stage of the XENON experiment \cite{xenon1t}.
While the assumed energy threshold of 2 keV might be optimistically low, our analysis is not very sensitive to this threshold, as we do not focus on particularly low WIMP masses. It would be interesting to see whether other future projects for direct detection, like an upscaled CoGeNT \cite{cogent} with keV threshold would increase the potential for direct detection appreciably. We leave this for future work, just noting the irreducible neutrino background that eventually will set in -- see the discussion at the end of this Section.

Fig.~\ref{figDD} certainly illustrates the importance of direct detection experiments, which not only start to actually exclude viable dark matter models but also show an encouraging potential to considerably strengthen these limits in the next decades; however, it also makes  clear that using direct detection methods only, success is by no means guaranteed, even in the ``standard'' case of having an MSSM template neutralino as the dark matter. 
Even upcoming data from CERN's LHC may in unfavourable cases not add too much discovery potential of the neutralino, especially if it is heavier than a few hundred GeV. (On the other hand, if the Higgs particle is discovered with a mass below some 130 GeV, the supersymmetric scenario would of course gain strength considerably.)

One should also keep in mind that the supersymmetric parameter space is multidimensional and extremely complex. Only recently have methods been developed which allow a more complete scan (see e.g.\ \cite{trotta,yashar} for examples on mSUGRA models). Therefore it is not excluded that the ranges in the $\sigma_{SI}-m_\chi$ plane are actually even more extended. In fact,
in Ref.~\cite{Mandic:2000} it was shown that for generic MSSM models essentially arbitrarily low cross sections are possible. This can, e.g., happen when the coupling to the heavy CP-even Higgs boson $H_1^0$ vanishes, or (more likely) when a cancellation occurs between the scattering diagrams with $H_1^0$ and (the light CP-even) $H_2^0$ exchanges, which can happen when $\mu<0$. Exchanges of squarks can also contribute, but their contributions are small if the squarks are heavy enough, or they can also be cancelled out by the Higgs exchange diagrams.
One should also note that in a frequentist approach, the parameter volume is actually irrelevant (and in particular, one should avoid associating any probability to individual points in the scans).
We emphasize that our scans of parameter space have {\em not} been designed to find models of particularly low (or high) direct detection cross sections. It may well be that some specific models would, e.g, be lifted to higher direct detection cross sections by radiative corrections or other effects. However, the low cross section regions of parameter space would then be filled in by other models due to similar corrections.

Of course, it would also be feasible to build even larger direct detection experiments. A 5--10 ton Xenon detector would be able to reach scattering cross sections down to about $10^{-12}$ pb \cite{laura-private}. However, 
one should remember that eventually direct detection experiments will face an (almost) irreducible background coming from atmospheric neutrino interactions in the detector \cite{nu_bg}.
For $10^{-10}\,{\rm pb}\gtrsim\sigma_{SI} \gtrsim 10^{-12}\,$pb, basically only  neutrinos produced in the sun from $^8{\rm B}\rightarrow\,^2{\rm Be}^*+e^++\nu_e$ constitute a background to the WIMP-induced recoil rate; these could in principle either be eliminated by setting the threshold to 5 keV (for Xe) or 7 keV (for Ge)  -- which, however, would also cut away a considerable part of the signal, in particular for low-mass WIMPs -- or by performing a more demanding multi-component fit to both the DM and the neutrino part of the recoil spectra \cite{Strigari:2009bq}. For even smaller $\sigma_{SI}\lesssim 10^{-12}\,$pb, many more neutrino sources contribute to the background and, even though the shape of the energy recoil spectrum will be different for neutrinos than for WIMPs, it will certainly be a great challenge to go far below this limit with direct detection experiments.

\section{MSSM and indirect detection}

In order to assess whether indirect dark matter searches can reach parts of the parameter space that are not accessible by direct searches, we will in the following consider the same sample of MSSM and mSUGRA models as before. As a first approach to a thorough treatment of this question, we focus on gamma-ray searches; noting, however,  that it would certainly be interesting to extend this analysis by including other messengers as well.
Multi-wavelength photon searches seem particularly promising given that radio signals, e.g., may well be more constraining than gamma rays \cite {Regis:2008ij}, especially for very cuspy dark matter profiles. This will be left for future studies, however.

Targets for indirect dark matter searches in gamma rays include the galactic center, dwarf spheroidal galaxies, dark matter clumps, galaxy clusters and the galactic as well as the extragalactic gamma-ray background -- out of which the first two are maybe the most promising. While the galactic center is expected to be the single-most luminous source of dark matter annihilations in the sky, it is also a region of very high astrophysical activity, and the signal has therefore to be searched for against  a large background that is difficult to understand in detail. The annihilation flux expected from dwarf galaxies, on the other hand, is at least one order of magnitude smaller than from the galactic center; these objects, however, are completely dominated by dark matter, with mass-to-light ratios of up to about 1000, so no considerable astrophysical background in gamma rays is expected. It seems that by in addition stacking data from many different dwarf galaxies, using the important fact that the energy spectrum from annihilating dark matter has to be the same for all sources, an important improvement of sensitivity can be obtained \cite{Gaarde:2010}. As this is still in a preliminary stage, we will not present results of this method here but only point out that this would further increase the reach of indirect detection by gamma-rays.
In the following, we will focus on the galactic center and a generic dwarf galaxy (like Draco or Willman 1), as they serve as illustrating examples of other possible cases that are limited by backgrounds and statistics, respectively.

\begin{table}[t]
\centering
\begin{tabular}{l|cll}
     {\bf Target}  &\hspace*{3ex}&& $J$ \\ 
\hline
\hline
Draco && $0.19$ & -- ~~ $3.0$\\
Willman 1 && $0.67$ & -- ~~ $60$\\
Ursa Minor && $0.061$ & --  ~~ $20$ \\
Sagittarius && $0.088$ & -- ~~ $183$\\
Segue 1 && $0.23$ & -- ~~ $22$ \\
\hline
GC, isothermal sphere &&  $0.013$ & ~~~ ($1.3$) \\
GC, NFW &&  $16.9$ & ~~~ ($167$)\\
GC, NFW+adiab. contr.  && $3.1\cdot10^4$ & ~~~ ($5.3\cdot10^4$)\\
\end{tabular}
\caption{\label{tab_J} The range of possible values for the astrophysical factor $J$, defined in Eq.~(\ref{BJdef}). For dSphs, kinematical data is used to determine the allowed range at 90\% \emph{c.l.} \cite{Essig:2009jx,Essig:2010em}; the additional boost due to substructures may be as large as $B\sim10$ -- $100$ \cite{Strigari:2006rd}. For the galactic center, the value for three benchmark dark matter profiles that we will refer to is shown (as implemented in \ds\ \cite{darksusy}), integrating over a region of $\Delta\Omega=10^{-6}$ $\left(10^{-4}\right)$.}
\end{table}

The expected gamma-ray flux from a cone with solid angle $\Delta\Omega$ observed at earth can be written as 
\be
 \label {eq_gamma}
  \frac{d\Phi_\gamma}{dE}=  \frac{\langle\sigma v\rangle}{2m_\chi^2}\sum_f \frac{dN^f_\gamma}{dE}\times  \frac{B}{4\pi}\int_{\Delta\Omega}\! d\Omega\int_{l.o.s.} d\lambda\, \langle\rho\rangle^2\,,
\ee
where $\langle\sigma v\rangle$ is the total annihilation rate, $N_\gamma^f$ the number of photons produced in annihilation channel $f$, $\rho$ the dark matter density (of which $\langle\rho\rangle$ is the smooth component only) and $\lambda$ is the line-of-sight distance in the direction of observation; for a realistic modeling, this expression has of course to be convolved with the energy and angular resolution of the detector (which we do take into account here). The enhancement of the signal due to the effect of dark matter substructure is encoded in what is often referred to as the "boost factor"
\be
  B\equiv\frac{\int_{\Delta\Omega}\! d\Omega\int_{l.o.s.} d\lambda\, \rho^2}{\int_{\Delta\Omega}\! d\Omega\int_{l.o.s.} d\lambda\, \langle\rho\rangle^2}\,.
\ee
Unfortunately, it is rather difficult to make any definitive statement about its size since it depends critically on the profile and distribution of subhalos much smaller in size than what can currently be reached in simulations. Extrapolating the results of such simulations down to the smallest subhalos (with a mass of $10^{-11}M_\odot$ to $10^{-3}M_\odot$, depending on the particle nature of dark matter \cite{mh}), and including the impact of sub-subhalos, it could be larger than 10 for the galactic center \cite{vl_I}, though most recent simulations tend to find only $\mathcal{O}(1)$ effects \cite{aquarius,vl_II}. 

The above expression for the gamma-ray flux can nicely be separated in one part that depends only on particle physics and one that depends on the distribution of dark matter, which is essentially determined by astrophysics and unfortunately not very well constrained by observations. In a dimensionless form, the latter reads $BJ$ with
\bea
 \label{BJdef}
  J&\equiv& \frac{1}{4\pi}\int_{\Delta\Omega}\! d\Omega\int_{l.o.s.} d\lambda\, \langle\rho\rangle^2
                        {\Big/}(10^{11}M_\odot^{2}{\rm kpc}^{-5} )\nonumber\\
       &\simeq& \frac{1}{4\pi D^2}\int d^3x\, \langle\rho\rangle^2
                        {\Big/}(10^{11}M_\odot^{2}{\rm kpc}^{-5} )\,,
\eea
where the last expression is valid for point-like sources at distance $D$, or sources that are (almost) contained in $\Delta\Omega$,  and the integral is to be performed over the spatial extent of the source.
Since it is so difficult to directly measure the dark matter profile, one basically has to rely on the result of high-resolution $N$-body simulations which support an NFW profile with its characteristic $\rho(r)\propto r^{-1}$ cuspy behaviour in the innermost part of the halo. However, these simulations do not take into account the effect of baryons in a fully self-consistent way. Whether, and to what extent, the existence of baryons leads to a further steepening of the dark matter profile due to adiabatic contraction \cite{Blumenthal:1985qy} is a matter of current debate. Observations of rotation curves, on the other hand, are sometimes used to argue for a shallower, cored profile \cite{Burkert:1995yz,Gentile:2004tb}. Overall, this leads to a rather large uncertainty in the total annihilation flux from the galactic center, though one may argue that a standard NFW profile is a reasonable bet (which still may considerably underestimate the actual flux). In the case of dwarfs, which are dark-matter dominated systems, this assumption is probably well satisfied; in fact, one can use kinematical data from the stars to constrain the astrophysical factor rather well. Table \ref{tab_J} summarizes the allowed range of $J$ for the galactic center and some selected dSphs.

\section{FERMI, CTA and DMA - the Dark Matter Array}

\begin{figure*}[ht!]
  \begin{minipage}[t]{0.49\textwidth}
      \centering
  \includegraphics[width=0.95\columnwidth] {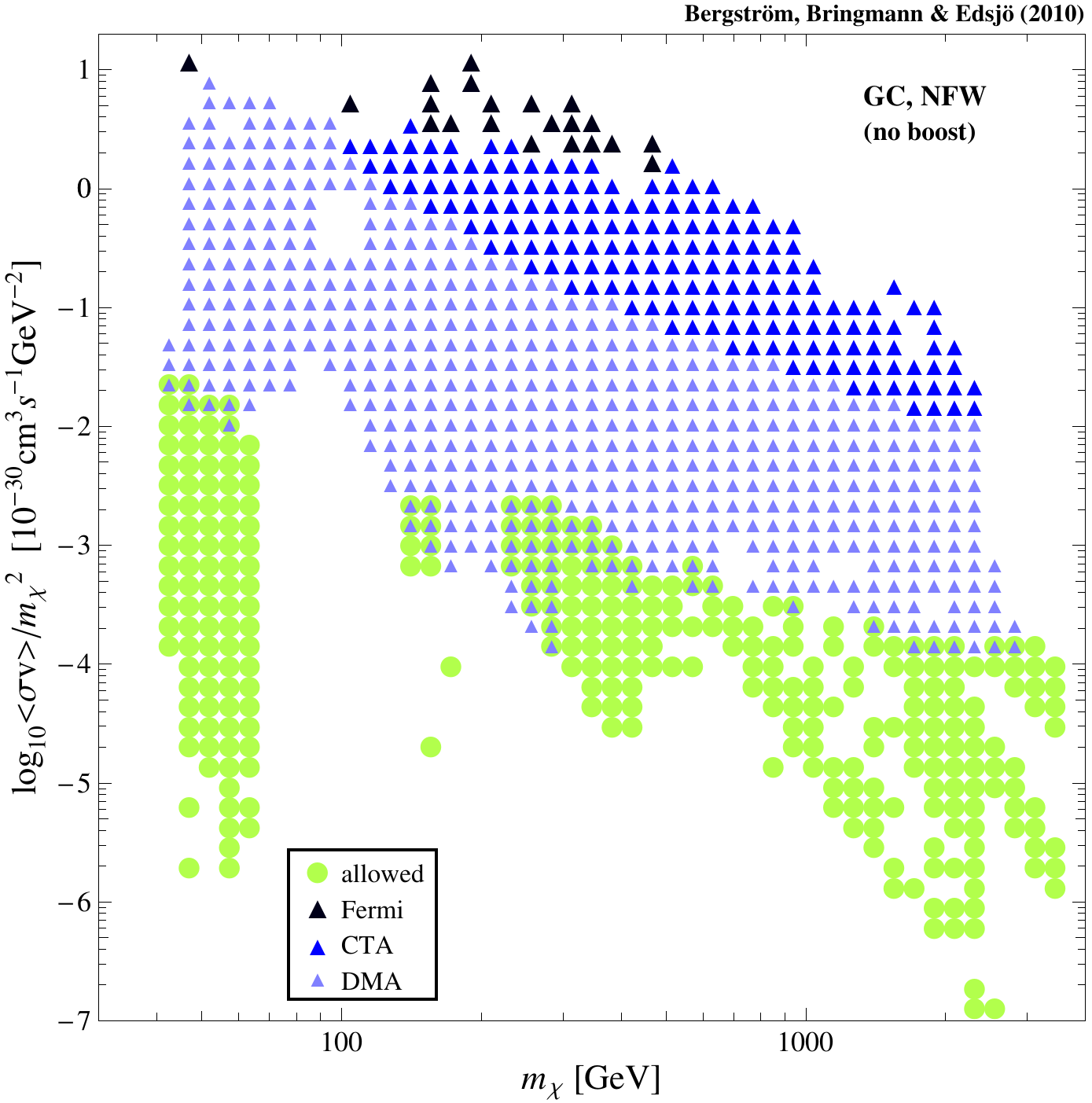}\\
   \end{minipage}
  \begin{minipage}[t]{0.02\textwidth}
   \end{minipage}
  \begin{minipage}[t]{0.49\textwidth}
      \centering   
  \includegraphics[width=0.95\columnwidth] {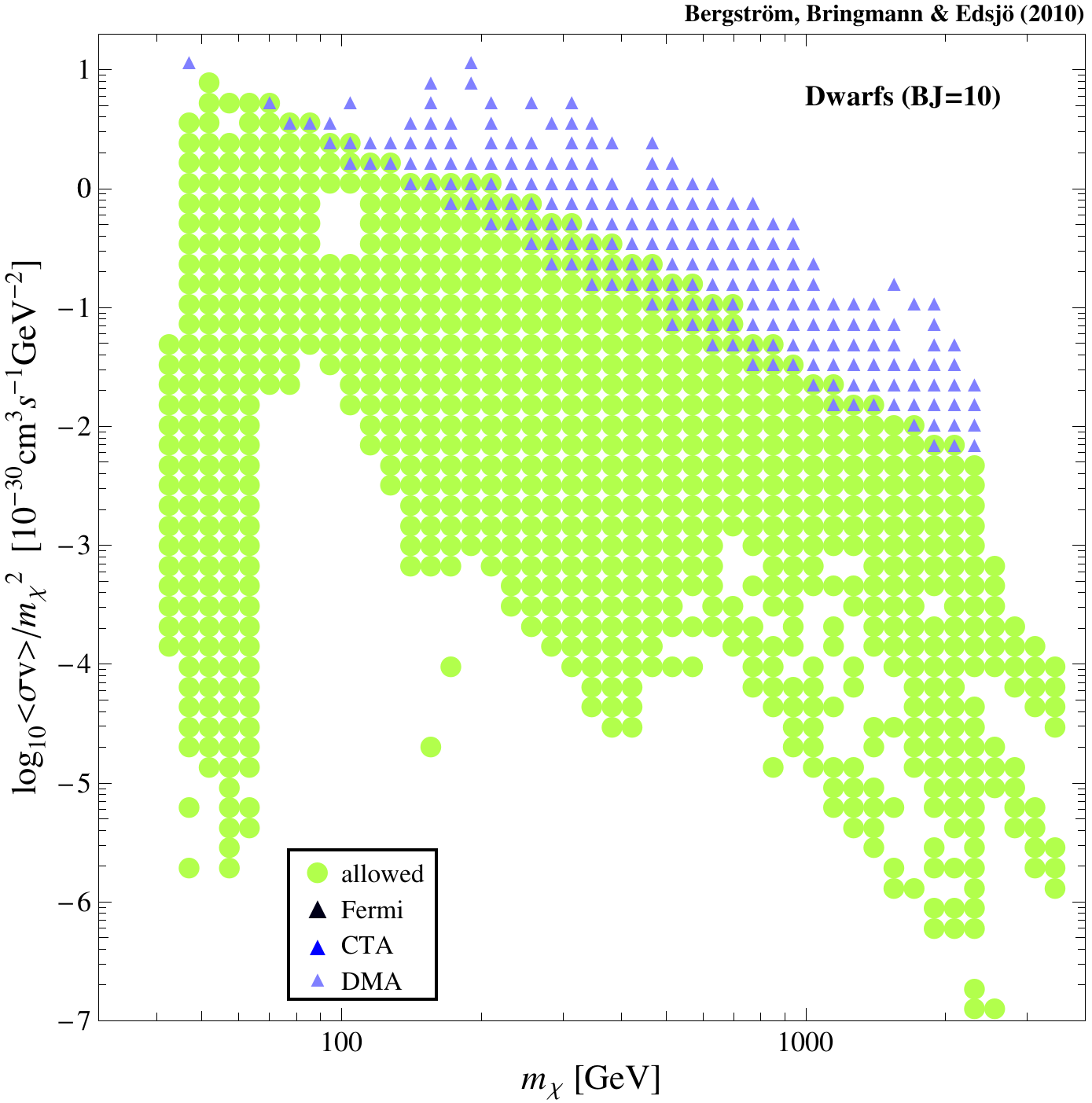}\\
   \end{minipage}  
{
\caption{The same set of MSSM and mSUGRA models as in Fig.~\ref{figDD}, this time plotted as the annihilation cross section (divided by $m_\chi^2$) versus the neutralino mass. Also shown are the projected limits from the gamma-ray experiments discussed in more detail in the text.}
\label{fig_indir}}
\end{figure*}

For what follows, we will focus on two examples of a space-based and a ground-based gamma-ray experiment that are considered to be milestones for the future of gamma-ray astronomy, 
 the currently operating Fermi satellite and the Cherenkov Telescope Array (CTA). The latter is currently still in the planning phase, but will  greatly exceed both sensitivity and energy range of the current generation of large ACTs like HESS \cite{hess}, MAGIC \cite{magic} and VERITAS \cite{Holder:2008ux}, aiming to reach milli-crab sensitivity in the $\sim100\,$GeV to $1\,$TeV range which is of particular interest to dark matter searches.

For our analysis of the Fermi discovery potential, we follow the approach of \cite{prelaunch} and consider 20 logarithmic bins in the $1-300\,$GeV energy range. Performance details like sensitivity, angular resolution, energy resolution and effective area we adopt from the LAT homepage \cite{LAT}. We assume an observation time of five years, roughly corresponding  to one full year of data taking. For CTA, we assume an energy threshold of $40\,$GeV and use the same logarithmic bin size as for Fermi up to the few TeV range. Still being in the planning stage, it is of course more difficult to make definitive statements about the final design; for definiteness, we will here use the projected sensitivity curves of \cite{CTA}, assume an angular resolution of 0.02$^\circ$ and, as usual for CTAs, an exposure time of 50 h. The effective area is taken from \cite{arribas}, with about $3\,$km$^2$ at $5\,$TeV, $1\,$km$^2$ at $100\,$GeV and a fall-off for lower energies down to $\sim0.1\,$km$^2$ at $40\,$GeV.

\begin{table}
  \begin{tabular}{ l || c | c | c}
    & {\bf Fermi} & {\bf CTA} & {\bf DMA}\\
    \hline\hline
    Energy resolution & $0.1$ & $0.1$ & $0.1$\\ 
    \hline
    Angular resolution [sr] & $10^{-4}$ & $4\cdot 10^{-7}$ & $4\cdot 10^{-7}$
     \vspace{-1ex}\\
     &&& {\scriptsize($10^{-5}$ for $E_\gamma<40\,$GeV)}\\ 
    \hline
    Energy threshold [GeV]  & $1$ & $40$ & $10$\\ 
    \hline
    Effective Area [m$^2$]& $0.7$ & $10^6$ & $10^7$\\
    \hline
    Observation time [h] & $10^4$ & $50$ & $5000$\\
  \end{tabular}
  \caption{\label{tab_exp}Comparison of performance details of Fermi, CTA and DMA as used in this study. Note that most of the above values are approximate and energy dependent; see text for the details of the actual implementation.}
\end{table}

For the case of the galactic center, we assume that the angular resolution of upcoming ACTs will be sufficient to discriminate sources like the HESS source very close to, but probably off-set from the center and unrelated to dark matter \cite{hess_GC}. For the diffuse background, we will use the model developed by the Fermi LAT group that describes the so far available data very well \cite{digel}. Though this background model is not yet officially released and in its final stage, this approach is more than sufficient for our purpose, keeping in mind that the understanding of the background will further improve with the advent of more data and that additional angular information about the annihilation signal could be obtained by looking slightly away from the galactic center, thus further improving the prospects for discriminating the dark matter signal from the background component. In what follows, we calculate the significance in each energy bin; in order to claim that a gamma-ray experiment is able to rule out a given dark matter model, we
 require that the largest significance computed in this way is at least at the $5\sigma$ level, i.e.~$S/\sqrt{S+B}>5$.\footnote{
Note that this rather simple approach is of course only intended to put possible \emph{constraints} on dark matter models and this is how the results in the next section concerning the galactic center should be interpreted. For the successful claim of the \emph{detection} of a dark matter signal, a much more sophisticated analysis would be needed; since the understanding of the background flux from the galactic center region is currently not yet well enough understood for this purpose, we leave this for a future investigation. We do not expect, however, that the resulting limits would change by more than a factor of a few.
}
 For the case of objects limited by statistics, like dSphs, we simply demand that the signal in at least one of the bins, properly smoothed by the energy resolution of the experiment, consists of at least 5 photons and is larger than the integrated sensitivity in that bin.

In order to assess the real potential of indirect searches we introduce here the idea of a \emph{dedicated} dark matter experiment. For concreteness, we focus on a specific ground-based detector, the Dark Matter Array, which we propose to essentially have a CTA-like setup but  optimized for dark matter searches. Let us now estimate how much such a design could improve the sensitivity of CTA, for  a realistic construction in the foreseeable future. First of all, let us note that an energy threshold of down to 5 GeV seems to be possible with current technology, if the array is moved to high altitudes above sea-level \cite{5at5}. The effective area could more or less straight-forwardly be increased by one order of magnitude and the observation time for dark matter targets by almost two orders of magnitude; in total, this would roughly result in an improvement of the integrated sensitivity by two orders of magnitude as compared to CTA. To give a rough idea of the potential of such an instrument, we will assume in the following, for definiteness,  a threshold of 10 GeV, an effective area 10 times the one of CTA and 5000 h observation time. Of course, these values should just taken to be indications of what the final design could look like, the details having to be worked out in more dedicated future studies. In Tab.~\ref{tab_exp}, we indicate how the performance details of DMA compare to those of CTA and Fermi.

To summarize, Fig.~\ref{fig_indir} shows the reach of the gamma-ray experiments discussed here in the plane $\langle\sigma v\rangle/m_\chi^2$ vs. the dark matter mass $m_\chi$, the first quantity being directly proportional to the expected signal as given in Eq.~(\ref{eq_gamma}). One can clearly see that while CTA would be able to assess a considerably larger class of models than Fermi, the reach of DMA would extend even much further into the underlying parameter space, illustrating nicely the potential of using a dedicated approach to indirect searches as proposed here.

\begin{figure}[t!]
 \includegraphics[width=0.95\columnwidth]{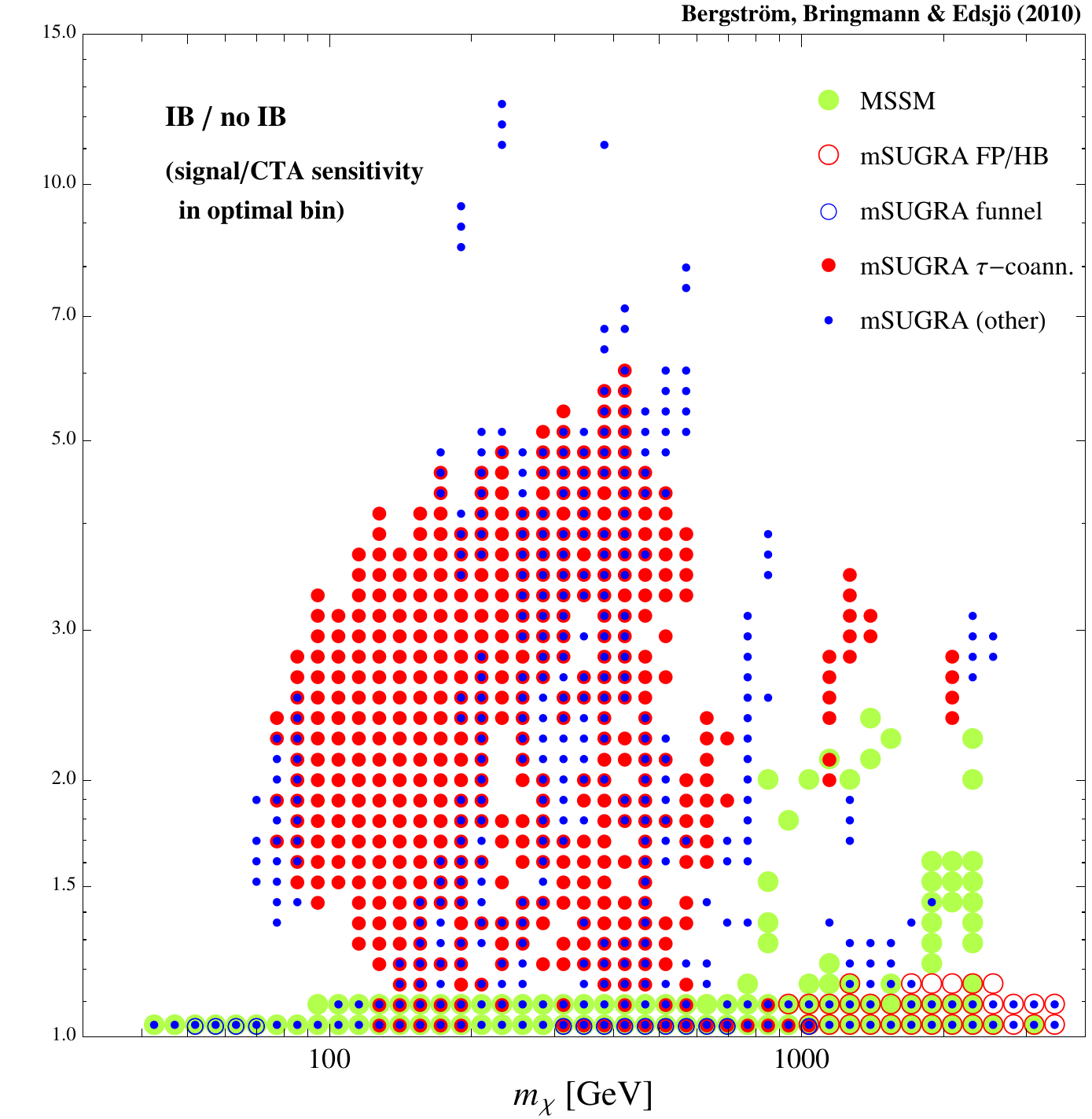}
\caption{This figure shows how much the detectability of a given model, defined here as the largest ratio of annihilation signal to CTA sensitivity, for all energy bins, is enhanced by including IB contributions.}
\label{fig_IB}
\end{figure}

\begin{figure*}[t!]
  \begin{minipage}[t]{0.49\textwidth}
      \centering
  \includegraphics[width=0.95\columnwidth] {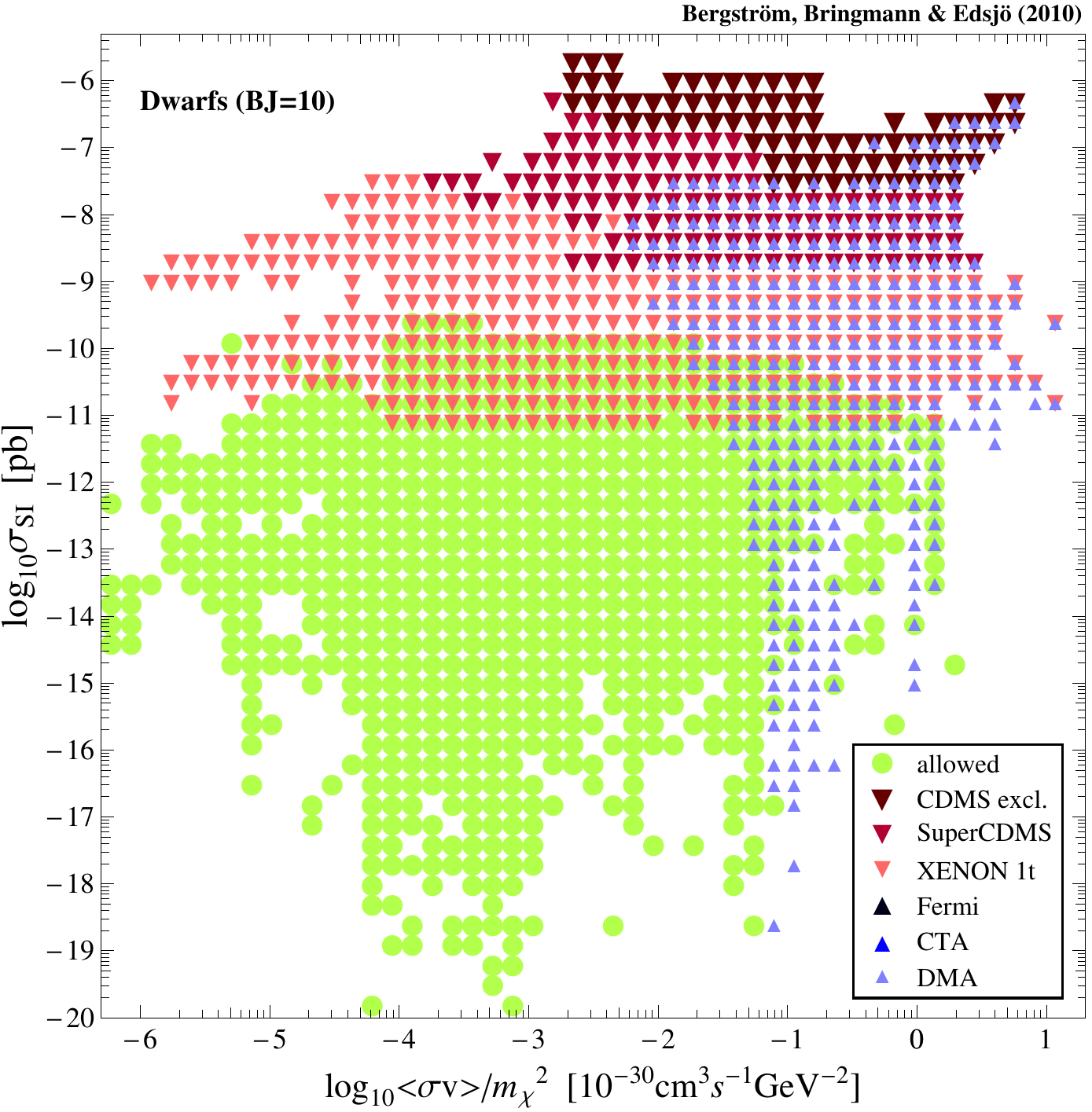}\\
   \end{minipage}
  \begin{minipage}[t]{0.02\textwidth}
   \end{minipage}
  \begin{minipage}[t]{0.49\textwidth}
      \centering   
  \includegraphics[width=0.95\columnwidth] {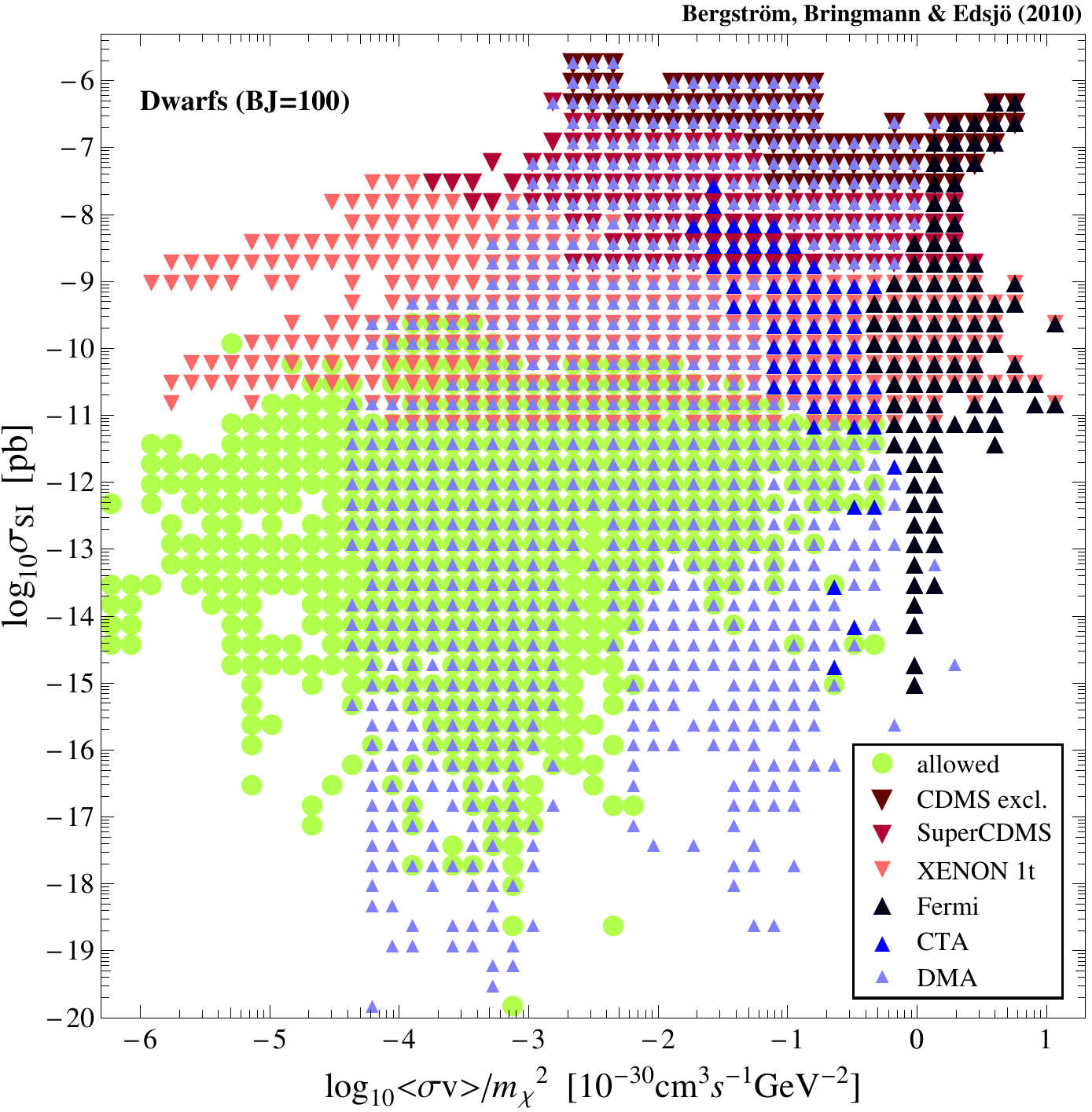}\\
   \end{minipage}  
{
\caption{The reach of direct detection experiments compared to that of indirect searches for dark matter in dwarf spheroidal galaxies. The left panel shows the situation for a rather typical dark matter distribution in these objects (taking into account the effect of substructures), while the right panel adopts rather favourable astrophysical assumptions. See text and Tab.~\ref{tab_J} for further details.}
\label{fig_dSph}}
\end{figure*}

In passing, we note that the contribution from IB is, indeed, quite important to take into account in this type of studies -- especially since it dominates the annihilation signal at high energies where ACTs are most sensitive. To illustrate this point, let us consider the ratio of expected signal and CTA sensitivity, which is of direct relevance for signal-dominated searches like in the case of dwarf galaxies. In Fig.~\ref{fig_IB}, we plot the enhancement of this quantity that is obtained by including IB effects as opposed to taking into account secondary and line photons only. As can be seen, the effect is most important for mSUGRA models in the bulk and stau coannihilation region, but it is certainly not negligible also at TeV neutralino masses like in the focus point/hyperbolic branch region. In fact, the chance to detect a given model is increased by up to an order of magnitude, in agreement with what was found in \cite{Bringmann:2008kj, Cannoni:2010my}. We take the opportunity to comment that while this is an important effect, the spectral signatures connected to these contributions are probably even more important as they would allow a much easier discrimination from astrophysical backgrounds than secondary photons alone.

\section{Direct versus indirect searches}

\begin{figure*}[t!]
  \begin{minipage}[t]{0.49\textwidth}
      \centering
  \includegraphics[width=0.95\columnwidth] {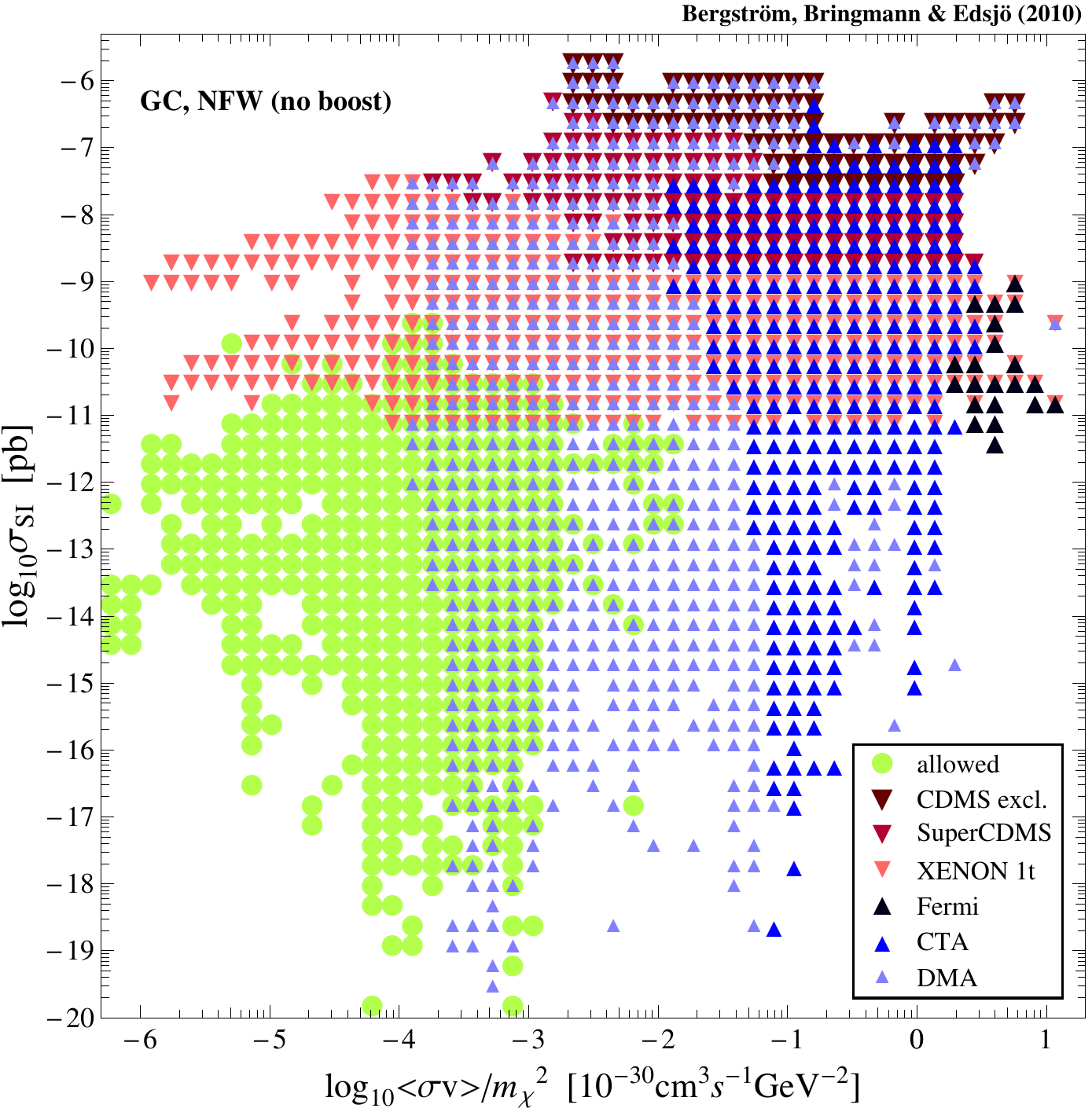}\\
   \end{minipage}
  \begin{minipage}[t]{0.02\textwidth}
   \end{minipage}
  \begin{minipage}[t]{0.49\textwidth}
      \centering   
  \includegraphics[width=0.95\columnwidth] {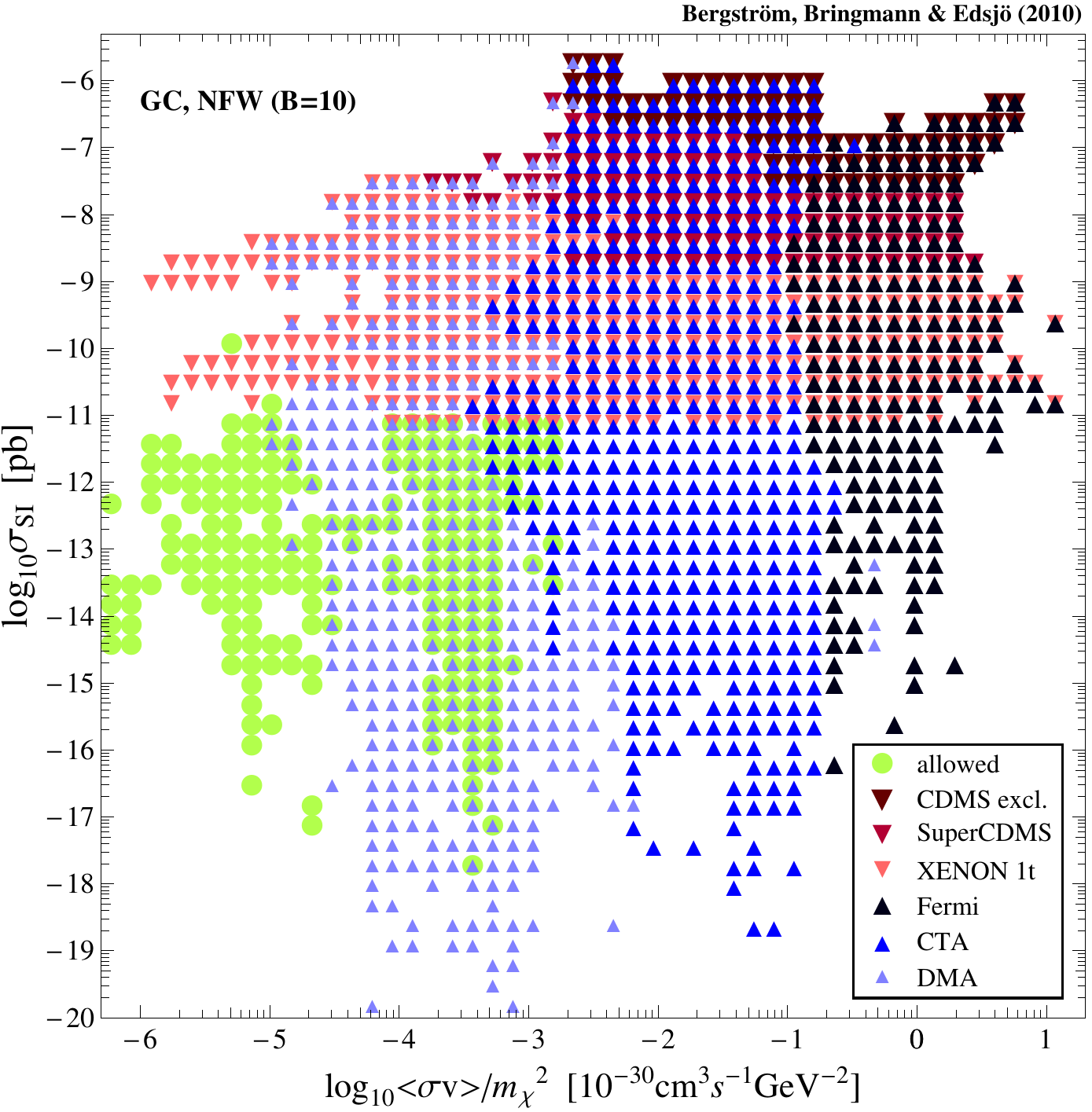}\\
   \end{minipage}  
  \begin{minipage}[t]{0.49\textwidth}
      \centering
  \includegraphics[width=0.95\columnwidth]{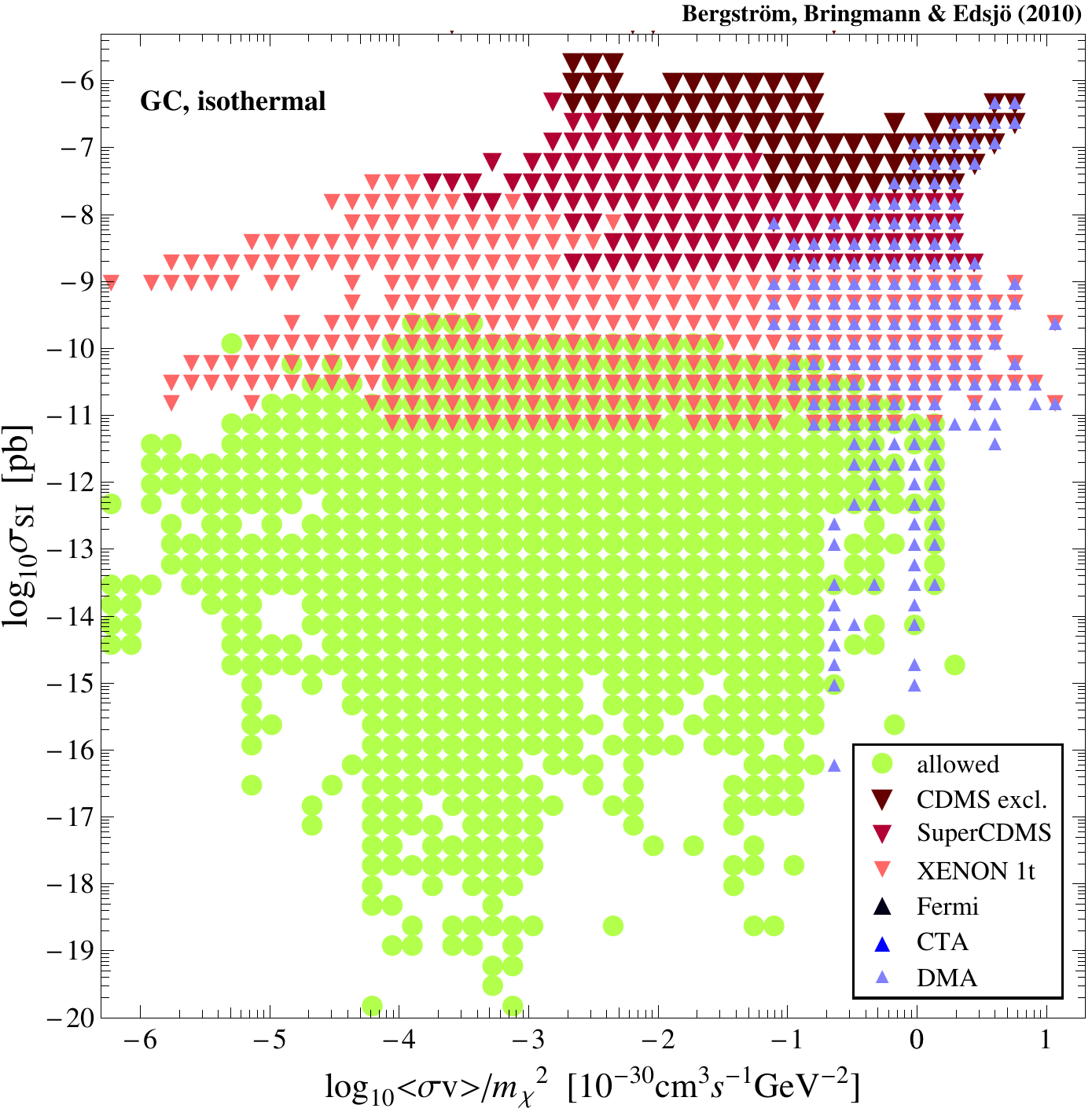}\\
   \end{minipage}
  \begin{minipage}[t]{0.02\textwidth}
   \end{minipage}
  \begin{minipage}[t]{0.49\textwidth}
      \centering   
  \includegraphics[width=0.95\columnwidth] {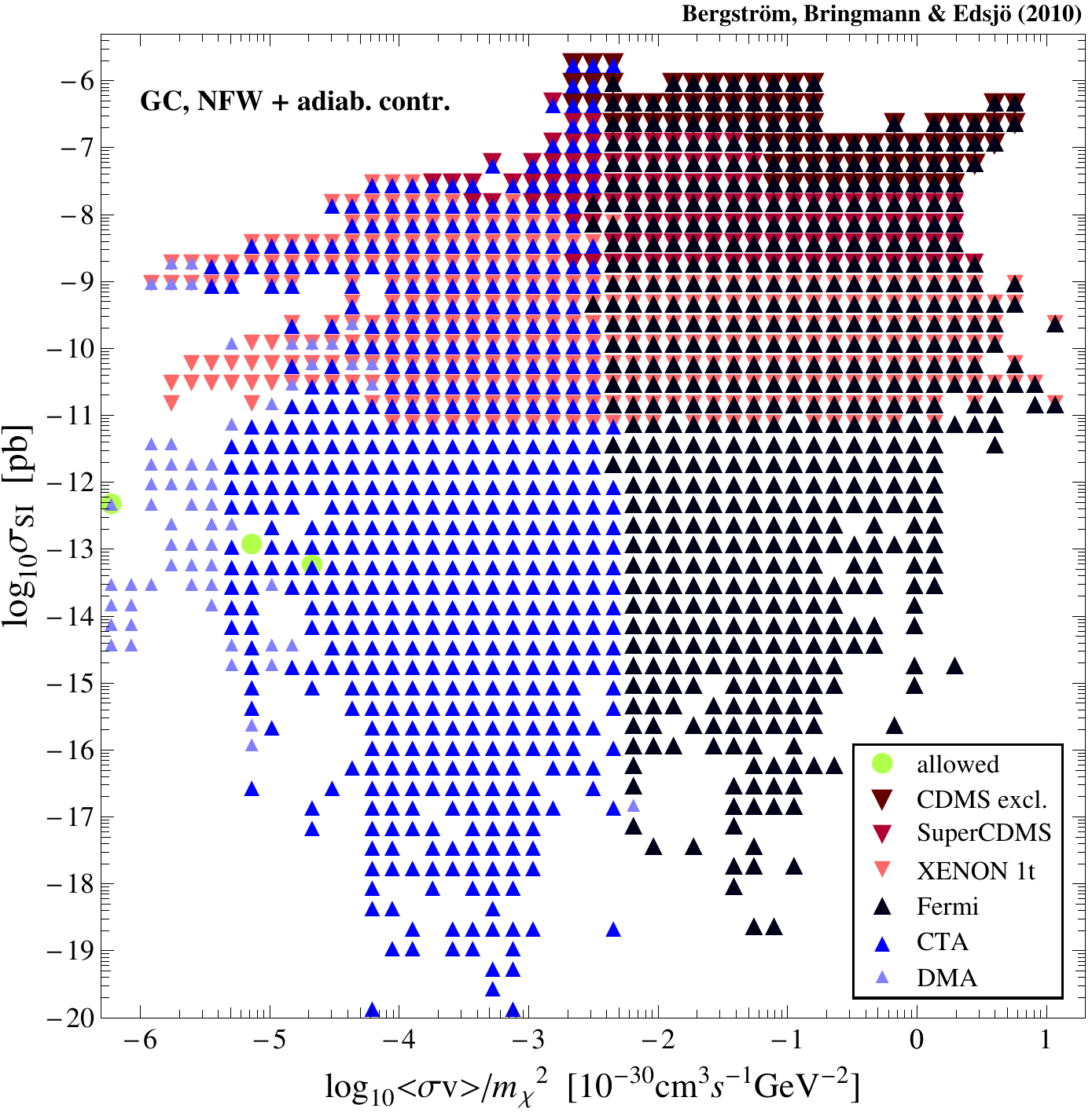}\\
   \end{minipage}
{
\caption{In the upper panels, the reach of direct and indirect dark matter detection experiments is shown for the NFW profile, which is supported by high-resolution $N$-body simulations, with and without the inclusion of a "boost-factor"  to account for the inhomogeneous distribution of dark matter. In the lower panels, for comparison, the same is shown for the extreme cases of a very shallow, cored isothermal sphere profile as well as the highly cuspy NFW profile after adiabatic contraction.}
\label{fig_gc}}
\end{figure*}

In order to emphasize the complementary nature of indirect and direct dark matter searches, we 
now combine the information contained in Figs.~\ref{figDD} and \ref{fig_indir} and plot our results in terms of  $\sigma_{SI}$ vs. $\langle\sigma v\rangle/m_\chi^2$. In Fig.~\ref{fig_dSph}, this is done for the case of  dwarf spheroidal galaxies. 
 The corresponding result for the galactic center is shown in Fig.~\ref{fig_gc},  for various assumptions about the dark matter halo profile.

As can be seen from these figures, dark matter indirect detection rates  for gamma-ray experiments are actually only very weakly correlated with direct detection rates. This means that
even in the (modestly) conservative cases of $BJ=10$ for dwarfs, or the NFW profile without boost factor for the galactic center, DMA permits a deep dive into parameter space, corresponding to direct detection rates way below what even a 10 t Xenon detector would be able to probe. For slightly more favorable dark matter profiles, or non-negligible contributions from substructures, a very large region of parameter space indeed would be covered. Of particular interest is of course also the upper right region in each of these plots, where the results from both indirect and direct searches will provide independent input to determine the nature of dark matter.

\begin{figure}[t!]
 \includegraphics[width=0.95\columnwidth]{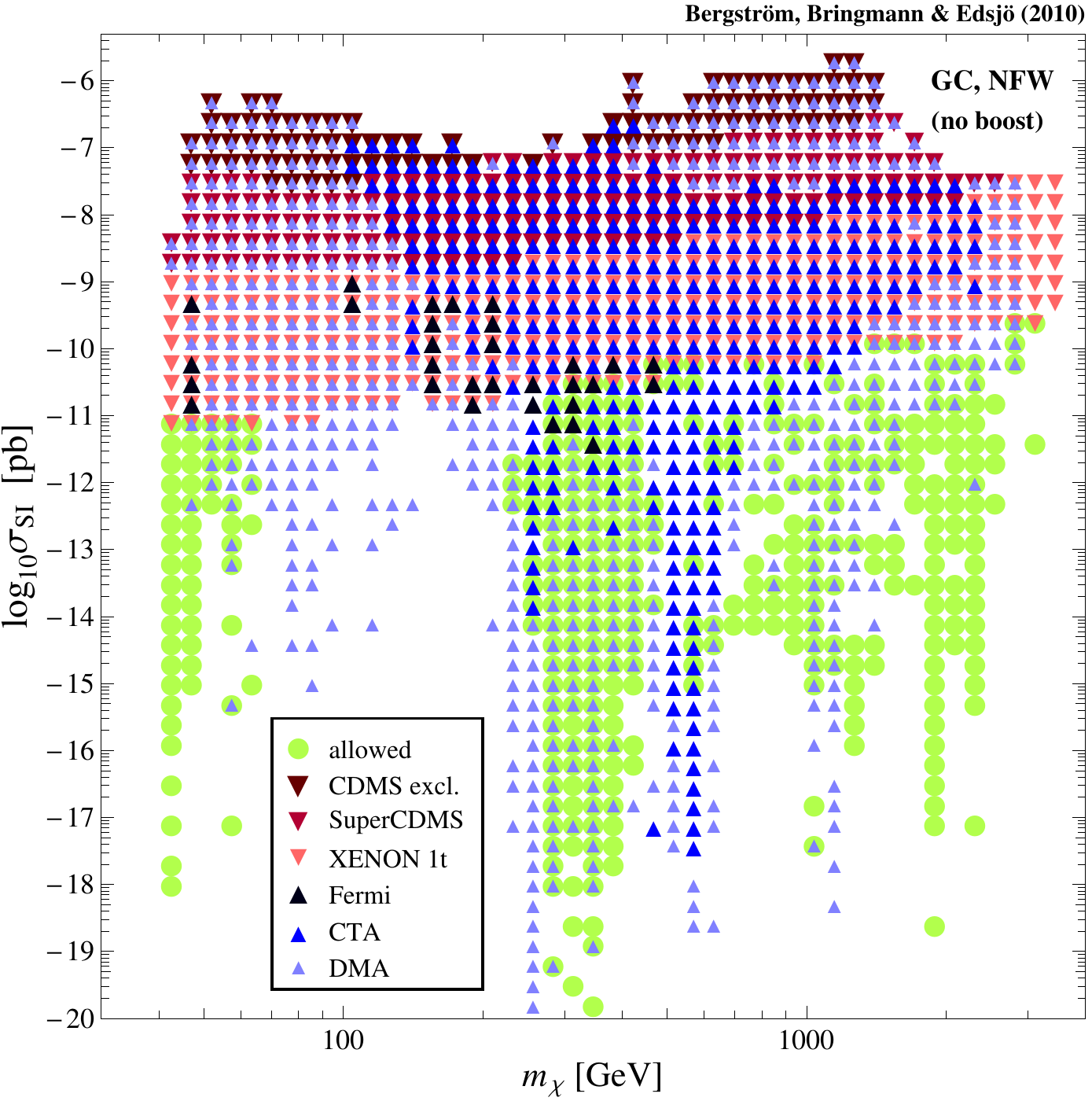}
\caption{ The models shown in the original plane of spin independent cross section versus neutralino mass, using the galactic center as target and assuming an NFW profile.}
\label{fig_sigsipmxindir}
\end{figure}

In order to develop a better understanding of which dark matter properties are easier to probe by direct and indirect searches, respectively, let us now revisit the original plane of spin independent cross section versus neutralino mass that was already shown in Fig.~\ref{figDD}; in Fig.~\ref {fig_sigsipmxindir}, we show the same plane, but also include the reach of indirect searches aiming at the galactic center (assuming an NFW profile). 
As can be seen, 
models with low $\sigma_{\rm SI}$ and very low masses will be difficult to access even with the next generation of indirect dark matter experiments; to probe as many models as possible in this region, it will, indeed, be very important to have an energy threshold as low as possible. One should note that in this region LHC will have a good potential of discovery.

\begin{figure}[t!]
 \includegraphics[width=0.95\columnwidth]{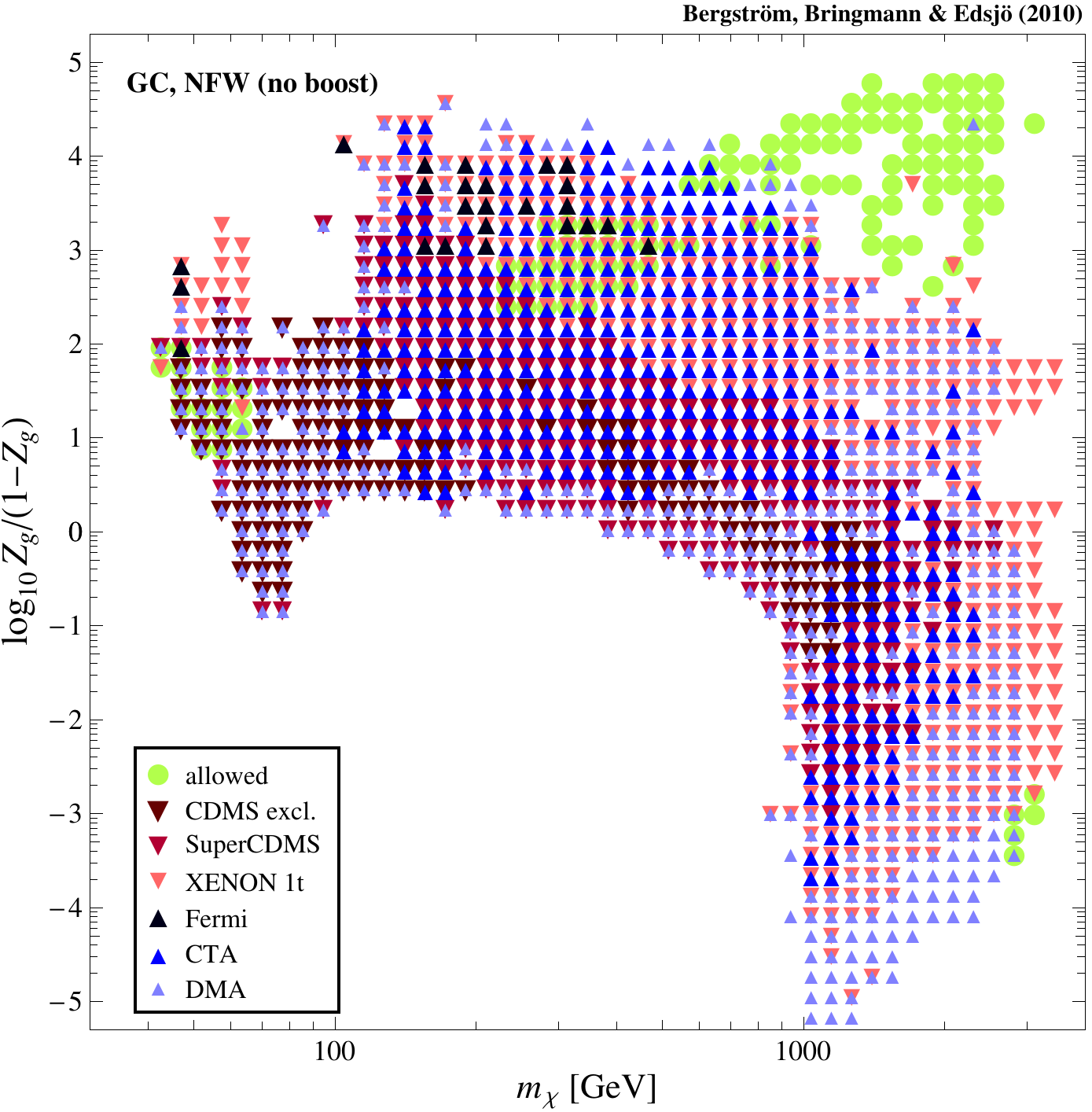}
\caption{Same as previous figure, but now plotting the gaugino fraction versus the neutralino mass. In the upper (lower) part, the neutralino is an almost pure gaugino (Higgsino). The middle part, particularly well suited for direct searches corresponds to the neutralino being a mixed state.}
 \label{fig_cgnfwvsmchi}
\end{figure}

Also some models with large masses seem difficult to be probed with indirect methods. In order to characterize better those models, we plot in Fig.~\ref{fig_cgnfwvsmchi} the gaugino fraction of the neutralino vs. its mass; apparently, models that cannot be accessed by neither direct nor indirect means are mostly rather massive, almost pure gauginos. An experiment like DMA could, however, probe these models for more favorable astrophysical configurations than a smooth NFW profile. 
It should also be noted that Sommerfeld enhancements  \cite{hisano} are expected to enhance the annihilation rate considerably for massive neutralinos with a high Higgsino or Wino component (though not so much for the Binos that appear in the top right corner of Fig.~\ref{fig_cgnfwvsmchi}). This effect has not yet been taken into account here and would of course improve detectional prospects.

We have here focused on comparing spin-independet direct detection experiments with gamma rays and shown that there is a large complementarity between the two. Of course, there are also other ways to search for dark matter, e.g. charged cosmic rays (positrons, antiprotons and antideutrons), neutrinos from the Sun/Earth and at the LHC. The charged cosmic rays can be a fruitful route to take if the dark matter signal is large enough and has a striking spectral shape. However, as has become clear in the aftermath of the Pamela positron excess, the poorly understood astrophysical backgrounds (mostly from primary sources) are a limiting factor.

Neutrinos from the Sun, though, are rather complementary to the signals we have discussed here, as that signal depends both on the spin-independent and the spin-dependent scattering cross section. For the latter, direct detection experiments are not very sensitive, and e.g. IceCube already now puts much stronger limits on the spin-dependent scattering cross section than direct detection experiments do. The spin-dependent scattering cross section is also not very correlated with the spin-independet scattering cross section, or the annihilation cross section. Hence, there are models that IceCube would be sensitive to that would not be visible in either direct detection experiments or gamma rays (note that focussing on the complementarity between neutrino signals and direct detection rates alone may not be sufficient to effectively reduce uncertainties in the determination of dark matter parameters \cite{Serpico:2010ae}). 

Some of the new particles in our supersymmetric models could also be produced at the LHC and one can roughly say that if the lightest sparticle masses are below a few hundred GeV, they would be visible at the LHC. Hence, LHC will probe the models to the left in, e.g., Fig.~\ref{fig_cgnfwvsmchi}.
A more complete study of the complementarity of all these different signals will be left for future work.

\section{Conclusions}

While single-purpose experiments are a concept very familiar from, e.g., direct searches for dark matter, gamma-ray telescopes have so far been operating as multi-purpose instruments for the study of extreme astronomical objects, with dark matter often considered but a mere side-aspect of the general science goals.
In this article, we have shown that  {\em
dedicated} 
observations, targeting high energy gamma-rays, would considerably improve
the prospects of detecting and identifying
WIMP dark matter candidates. 
 
In particular, we have
shown that a large class of even the most natural WIMP candidates, those from
the MSSM, may have direct detection cross sections too small to conceivably
be detected in presently planned experiments for direct detection. 
In fact, for the low cross sections we have found here, the background becomes a 
serious principal limitation so that even much more ambitious direct detection experiments would probably not be able to probe all the models we have presented here.
Also,
the neutralinos may be too massive for detection at the LHC, leaving a unique, very
interesting window accessible only for searches in gamma rays . 

On the other hand, it has also to be pointed
out that parameters may be such that direct detection is more favourable.
There is finally the ideal situation with the possibility of seeing a
signal in at least two independent types of experiments - something that
may eventually be needed to make a convincing case that the dark matter
problem has been solved. On their own, the results we have presented here
may be most useful when determining exclusion limits for WIMP candidates. 

The somewhat ideal case of detector we have presented, the DMA, would
probably be built in stages, which is one of the advantages of the imaging
air Cherenkov array method. In fact, a non-negligible improvement of limits
from CTA would be possible by devoting more time to dark matter surveys
with that instrument. One could also investigate the possibility of using
the H.E.S.S., MAGIC and VERITAS arrays for this purpose, especially at the
time when the CTA is ready. 
Eventually, to extract the most benefit from
the gamma-ray indirect detection method, a dedicated stand-alone experiment
may be needed, however. Of course, various elements of optimization both of
such a detector and of the choice of plausible dark matter sources will be
necessary, as will a more careful estimate of the level of back- and
foregrounds and other systematic uncertainties. This we leave for future
studies, however.

\smallskip
\acknowledgments
We wish to thank L.~Baudis, G. Bertone, J. Conrad, W. Hofmann, M. Teshima and  J. Silk  for stimulating discussions.
L.B. and J.E. acknowledge support from the Swedish Research Council (VR). T.B. acknowledges support from the German Research Foundation (DFG) through the Emmy Noether fellowship grant BR 3954/1-1.
We are grateful to E.A. Baltz for letting us use his extensive data 
base of mSUGRA models.

\end{document}